\documentclass[%
preprint,
 amsmath,amssymb,
 aps,
]{revtex4-1}
\usepackage{graphicx}
\usepackage{dcolumn}
\usepackage{bm}
\usepackage[margin=0.5in]{geometry}
\usepackage{amsmath}
\begin{document}
\title{Spatial distributions of non-conservatively interacting particles}
\author{
Dino Osmanovi{\'{c}}
}
\affiliation{Center for the Physics of Living Systems, Department of Physics, Massachusetts Institute of Technology, Cambridge, MA 02139, USA}%
\email{dinoo@mit.edu}
\date{\today}
\begin{abstract}
Certain types of active systems can be treated as an equilibrium system with excess non-conservative forces driving some of the microscopic degrees of freedom. We derive results for how many particles interacting with each other with both conservative and non-conservative forces will behave. Treating non-conservative forces perturbatevily, we show how the probability distribution of the microscopic degrees of freedom is modified from the Boltzmann distribution. We compare the perturbative expansion to an exactly solvable non-conservative system. We then derive approximate forms of this distribution through analyzing the nature of our perturbations. Finally, we consider how the approximate forms for the microscopic distributions we have derived lead to different macroscopic states when coarse grained, and compare it qualitatively to simulation of non-conservatively interacting particles. In particular we note by introducing non-conservative interactions between particles we modify densities through extra terms which couple to surfaces.
\end{abstract}
\maketitle

\section{Introduction}

Non-equilibrium systems remain a frontier for physics. Indeed, even the term ``non-equilibrium" applies to a large number of observed systems, and it is not immediately apparent whether there would exist a single framework which would unify all these disparate phenomena. One particular class of non-equilibrium systems, deemed ``active matter", focuses on situations where the microscopic components are being driven through continuous injection of energy\cite{marchetti_2013}. Active matter has generated special interest as continual consumption of energy microscopically is one of the characteristic properties of life\cite{needleman_2017}.

Even as non-equilibrium describes a wide variety of things, active matter also applies to a very diverse variety of phenomena, existing from sub-cellular length scales to interactions among organisms.  Theoretical treatments of these systems often proceed from writing down a model of the particular active system in question, usually starting from considering the forces involved\cite{fodor_2018,cates_2019}. One may then either simulate the microscopic equations of motion, or coarse-grain (either with top down or bottom up approaches) the microscopic degrees of freedom into effective field treatments that capture large scale properties. These methods are usually able to be successfully compared to the real experimental realizations\cite{bricard_2015}. Whether there exists a more generic conceptual basis for such systems is still a topic of ongoing research\cite{takatori_2015,marinibettolomarconi_2015,speck_2016}.

One salient difference between an active and an equilibrium system is the existence of non-vanishing microscopic currents\cite{zhang_2012}. In equilibrium all currents vanish, however for an active system this does not necessarily hold\cite{battle_2016}. Indeed, continual consumption of energy by the microscopic components somewhat implies the existence of currents. Nevertheless, active matter can relax to a non-equilibrium steady state, where the relevant statistical properties of the system are no longer changing in time, though microscopic currents may still be present. However, while for an equilibrium system the steady state properties can be calculated from the Boltzmann distribution, which can relate the microscopic degrees of freedom to the macroscopic properties via the partition function, steady state properties of active matter cannot necessarily be calculated via this method, as they are fundamentally out of equilibrium.

We can interrogate more deeply why exactly this may be. In particular, when one writes a model for an active system it may not be immediately apparent why this microscopic description would describe a non-equilibrium system. Were one to perform a careful analysis of the equations, however, one should find that there are terms in the dynamics which would necessitate the continuous injection of energy in order to be a realistic descriptor of a physical system. The microscopic equations may, for example, break the fluctuation dissipation theorem, corresponding to the fact that there may be active noise driving the components\cite{dabelow_2019}.  Another option is that, after one has written a microscopic description, one may see that there exists effective currents in the deterministic force part of the time evolution, a simple example could be that one microscopic degree of freedom $x_1$ is affecting the time evolution of another microscopic degree of freedom $x_2$ but that $x_2$ does not affect the time evolution of $x_1$ (or does not affect it in a commensurate way). Such a dynamical scheme would violate Newton's third law and could only be maintained by external energy input. This simple loop would correspond to there being a \textit{non-conservative force} in the system. The presence of microscopic currents at steady state also implies the existence of effective non-conservative forces in some of the microscopic degrees of freedom. A more complicated example of non-conservative forces in active matter would be the microscopic descriptions of chemically active systems\cite{osmanovi_2019}.

The stationary state of a system where some of the forces are non-conservative can no longer be characterized as an equilibrium one, nor could the microscopic probability distribution be constructed from a Hamiltonian, as the forces in the system no longer arise as the gradient of a Hamiltonian. More broadly, the steady states of the system have to be understood from a consideration of the dynamic processes involved rather than from construction via the Boltzmann distribution. 

Inspired by these ideas, in this paper we consider more generally what impacts microscopic driving, implemented as microscopic non-conservative forces, can lead to when describing collections of particles. In section \ref{sec:analytic} we discuss what effect a non-conservative force perturbing an equilibrium system has in terms of modifications of the ordinary Boltzmann distribution. Subsequently in section \ref{sec:appr} we test these perturbation expansions for a linear system, where the microscopic probability distribution for a generic linear force field with conservative and non-conservative parts can be solved exactly using matrix methods. We then derive effective forms of the stationary distribution for arbitrary non-conservative forces. Following on from this in section \ref{sec:shape} we discuss how microscopic non-conservative forces lead to changes on macroscopic observables such as density, and demonstrate the qualitative validity of our approach against simulations of particles interacting under a non-conservative pair force. Finally, we discuss in more depth the physics that emerges from statistical ensembles of non-conservatively interacting particles in section \ref{sec:diss}.

\subsection{Non-conservative forces}
Before we discuss our main results, we will briefly review conservative and non-conservative forces, as these concepts will appear repeatedly throughout the rest of the text. The presence of a non-conservative forces imply that the system has some form of path dependence. A conservative force is one in which the the total work done around any closed loop is zero:
\begin{equation}
\oint \mathbf{F}_{c}(\mathbf{r}).\mathrm{d}\mathbf{r}=0
\end{equation}
Or alternatively, the total work done along a path depends only on the endpoints of the path. Non-conservative forces do not have this property:
\begin{equation}
\oint \mathbf{F}_{nc}(\mathbf{r}).\mathrm{d}\mathbf{r}\ne 0
\end{equation}
By the gradient theorem, a conservative vector field can be written as a gradient of a scalar potential:
\begin{equation}
\mathbf{F}_{c}(\mathbf{r})=\nabla f(\mathbf{r})
\end{equation}
However, non-conservative vector fields cannot be represented in this way. For vector fields in three dimensions, it is well known that any vector field (with proper conditions) can be decomposed into a gradient part and a curl part through the famous Helmholtz decomposition\cite{griffiths_2017}:
\begin{equation}
\mathbf{F}(\mathbf{r})=-\nabla \psi(\mathbf{r})+\nabla\times\mathbf{\alpha}(\mathbf{r})
\end{equation}
where $\psi$ is some scalar field and $\mathbf{\alpha}$ is a vector field. From now on we will use interchangeably the terminology that gradients are conservative and curls are non-conservative. Certain generalizations to the Helmholtz decomposition exist for vector fields in higher dimensions, through Hodge theory, with mathematical assumptions over the domain and that $F$ is sufficiently well behaved. In equilibrium systems, the interparticle forces arise from potentials, and are thus conservative. However, for non-equilibrium systems this will not be generically true. Additionally, a system which is being driven by non-conservative forces would require there to be some energy source in order to continue its drive.

 The existence of vector field decomposition is suggestive when considering some problems in active matter, as the non-conservative part of the forces could be the part that is necessary for driving the system out of equilibrium. These terms could either be imposed or identified by hand, or for more realistic forces, decomposition algorithms could identify which part of the force field is either conservative or non-conservative\cite{deriaz_2009}. Were one to note that the non-conservative part is zero, the system is either equilibrium with an effective Hamiltonian, or is being driven out of equilibrium by another mechanism.

While this framework is more familiar from electrodynamics, in the supplementary information (section S1) we show how some different example active matter systems can be cast in terms of the interplay of conservative and non-conservative forces.

\section{Particles with non-conservative forces} \label{sec:analytic}

We write down the generic time evolution equations for the microscopic degrees of freedom of a system of $N$ particles evolving under a force field inside a dissipative medium:
\begin{align}
\frac{\mathrm{d} \mathbf{r}}{\mathrm{d}t}&=\frac{\mathbf{p}}{m} \\
\frac{\mathrm{d} \mathbf{p}}{\mathrm{d}t}&=\mathbf{F}\left(\mathbf{r}\right)-\gamma \mathbf{p}+\xi(t)
\end{align}
where $\mathbf{r}$ is the vector of all the positions of particles in the system, and $\mathbf{p}$ is the vector of all the momenta in the system:
\begin{align}
&\mathbf{r}=\left(\mathbf{r}_1,\mathbf{r}_2,\hdots,\mathbf{r}_N\right) \\
&\mathbf{p}=\left(\mathbf{p}_1,\mathbf{p}_2,\hdots,\mathbf{p}_N\right)
\end{align}
We have employed the assumption for this model that the force field $\mathbf{F}$ depends on the position degrees of freedom of the system only $\mathbf{F}$. This assumption could in principle be relaxed in order to describe different classes of active systems where the forces depend on the momenta. We also assume that these particles exist in a media with a dissipative term $\gamma$ and a white noise term $\xi$ which has the property that $\langle \epsilon(t)\epsilon(t')\rangle =2 k_b T \gamma \delta(t-t')$.

We now restrict ourselves to analysis of the stationary distribution of systems with this time evolution. We further focus our attention on the case where the system is overdamped. We wish to obtain knowledge of the microscopic probability of as a function of the positions of all the particles $P(\mathbf{r})$. The corresponding Smoluchowski equation for the evolution of the probability density under these forces is given by\cite{risken_1991}:
\begin{equation} \label{eq:Smul}
\frac{\partial P(\mathbf{r},t)}{\partial t}=\nabla.\left(-\mathbf{F}\left(\mathbf{r}\right)P(\mathbf{r},t)+k_b T \nabla P(\mathbf{r},t)\right)
\end{equation}
Where we have introduced the $\nabla$ differential operator, which is given by $\nabla=\left(\frac{\partial }{\partial \mathbf{r}_1},\frac{\partial }{\partial \mathbf{r}_2},\hdots,\frac{\partial }{\partial \mathbf{r}_N}\right)$. As the probability must be greater than or equal to zero, we represent the probability density as $P(\mathbf{r})=\exp(-\phi(\mathbf{r}))$. The stationary state distribution is the one in which the time derivative of the probability is zero, in other words, the left hand side of equation \ref{eq:Smul} is equal to zero. From equation \ref{eq:Smul}, the stationary probability density in this system must obey the following relationship:
\begin{equation} \label{eq:fp}
\nabla.\left(\exp\left(-\phi(\mathbf{r})\right)\left(-\mathbf{F}(\mathbf{r})-k_b T  \nabla \phi(\mathbf{r})\right) \right)=0
\end{equation}
One can see trivially if the forces in the system arise from the gradient of a function (such as the Hamiltonian) $F(\mathbf{r})=-\nabla H(\mathbf{r})$ that the stationary probability distribution over microstates will be given by the Boltzmann distribution $P(\mathbf{r})=\exp(-H(\mathbf{r})/k_b T)$. For systems in which the forces do not arise as the gradient of a function, the solutions are not so trivial. We remind readers that we use the terms \textit{conservative} to describe forces which are expressible as the gradient of a scalar function, and \textit{non-conservative} to describe forces which cannot be expressed in this way. Following Risken and others\cite{risken_1991,wedemann_2016}, we imagine that the force can be split into two components:
\begin{equation}
F(\mathbf{r})=\mathbf{f}^{(c)}(\mathbf{r})+\mathbf{f}^{(a)}(\mathbf{r})
\end{equation}
Where the new conditions that have to be specified for the probability distribution to be stationary are given by:
\begin{align}
\mathbf{f}^{(c)}(\mathbf{r})&=-k_b T \nabla \phi(\mathbf{r})\label{eq:prob} \\
\nabla.\left(e^{-\phi(\mathbf{r})} \mathbf{f}^{(a)}(\mathbf{r})\right)&=0 \label{eq:nc}
\end{align}
However, this is only a redefinition of the problem equation \ref{eq:fp}. Were we to be able to find the proper splitting of the force fields we could in principle solve for the stationary distribution. We will now focus our attention on an example system given by forces:
\begin{equation}
\mathbf{F}(\mathbf{r})=-\nabla H_0(\mathbf{r})+\mathbf{m}(\mathbf{r})
\end{equation}
Where $H_0$ defines some equilibrium Hamiltonian, but we keep $\mathbf{m}(\mathbf{r})$ to be as general as possible, such that it may include conservative and non-conservative elements. We here note that the splitting operation defined in equations \ref{eq:prob} and \ref{eq:nc} is not the same as merely separating the conservative and non-conservative components of the force, as can be seen from equation \ref{eq:nc}, which will not generally be equal to zero even if $\nabla.\mathbf{m}(\mathbf{r})=0$. The challenge is therefore to find the splitting of the force field which satisfies these conditions. This is a difficult problem, but can be recast into the form of a differential equation by making the following addition, introducing the new scalar field $\chi$:
\begin{align}
f^{(c)}(\mathbf{r})&=-\nabla H_0(\mathbf{r}) - \nabla \chi(\mathbf{r}) \\
f^{(a)}(\mathbf{r})&=m(\mathbf{r})+\nabla \chi(\mathbf{r})
\end{align}
this trivially satisfies the conservative condition of the log probability equation \ref{eq:prob}. In order to find the probability distribution we are then left with satisfying the following differential equation for $\chi$ from equation \ref{eq:nc}:
\begin{equation}\label{eq:chi}
\nabla^2 \chi(\mathbf{r})-\beta(\nabla H_0(\mathbf{r})+\mathbf{m}(\mathbf{r})).\nabla \chi(\mathbf{r})-\beta(\nabla \chi(\mathbf{r}))^2=\beta \nabla H_0(\mathbf{r}).\mathbf{m}(\mathbf{r}) -\nabla.\mathbf{m}(\mathbf{r})
\end{equation}
where we have introduced the notation that $ k_b T = \beta^{-1}$. This is a $N$ dimensional non-linear partial differential equation in the unknown $\chi(\mathbf{r})$, which is by itself still a challenging proposition. However if we define the following substitution $\chi(\mathbf{r})=-\beta^{-1}\log\left(\mu(\mathbf{r})\right)$ we are left with the following linear equation in $\mu(\mathbf{r})$:
\begin{align}
&\left(\nabla^2-\mathbf{q_1}(\mathbf{r}).\nabla+q_2(\mathbf{r})\right)\mu(\mathbf{r})=0 \label{eq:mu} \\ 
&\mathbf{q_1}(\mathbf{r}) = \beta (\nabla H_0(\mathbf{r})+\mathbf{m}(\mathbf{r})) \\
&q_2(\mathbf{r}) = \beta^2 \nabla H_0(\mathbf{r}).\mathbf{m}(\mathbf{r})-\beta \nabla.\mathbf{m}(\mathbf{r}) \label{eq:q2}
\end{align}
Where only solutions with $\mu(\mathbf{r})>0$ are physical ($\chi(\mathbf{r})$ is real). We assume a perturbative expansion exists in the magnitude of the additional forces $\mathbf{m}$. Parameterizing the additional forces with a parameter $\epsilon$:
\begin{equation}
\mathbf{m}\to\epsilon \mathbf{m}
\end{equation}
and seeking solutions of $\mu(\mathbf{r})$ of the form $\mu(\mathbf{r}) \approx \sum_{n=0} \epsilon^n \mu_n(\mathbf{r})$.

Substitution of this expansion leads to the following differential equation for the $nth$ order perturbation:
\begin{equation} \label{eq:perturb}
\left(\nabla^2-\beta \nabla H_0(\mathbf{r}).\nabla\right)\mu_n(\mathbf{r})=(\beta \mathbf{m(r)}.\nabla-q_2(\mathbf{r}))\mu_{n-1}(\mathbf{r})
\end{equation}
Where the expression on the left is a linear operator acting on $\mu_n(\mathbf{r})$ and the term on the right is some forcing term. This series would be supplemented by the conditon that $\mu_0=\text{const}$. The equation to each individual $\mu_n$ is acted upon by the same linear operator for every $n$, the only differences are the \textit{source} terms on the right hand side of the equations. Therefore, were one to be able to find a Green's function of the linear operator acting on each individual $\mu_n$, given by the following equation:
\begin{equation} \label{eq:green}
\left(\nabla^2-\beta \mathbf{\nabla}H_0.\nabla\right)G(\mathbf{r},\mathbf{r'})=\delta(\mathbf{r-r'}) 
\end{equation}
with the appropriate boundary conditions that $G$ vanishes at infinity. Then one may express the entire series solution in terms of this function as:
\begin{align}
\mu_0(\mathbf{r}) &= \text{const} \\
\mu_1(\mathbf{r}) &= -\mu_0 \int G(\mathbf{r},\mathbf{r'})q_2(\mathbf{r'}) d\mathbf{r'} \\
\vdots \\
\mu_n(\mathbf{r}) &= \int G(\mathbf{r},\mathbf{r'}) (\beta \mathbf{m(r')}.\nabla'-q_2(\mathbf{r'}))\mu_{n-1}(\mathbf{r'})\mathrm{d}\mathbf{r'}
\end{align}
As the Green's function only depends on the \textit{equilibrium} properties of the system this representation shows the effect of non-conservative force fields in terms of integrals over effective source terms arising from the imposition of the non-conservative force. 

This series would be the steady state solution of the microscopic probability for a system evolving under a generic force field. However, the complexity of this representation is hardly less than the original equation, given the difficulty in evaluating the coordinate space Green's functions. We will revisit this question in a later section and for now also discuss the physical significance of the perturbations. For completeness, we note that the first order perturbation is given by:
\begin{equation} \label{eq:ssd}
P(\mathbf{r})\approx \exp\left(-\beta H_0(\mathbf{r})+\epsilon \beta \int \mathrm{d}\mathbf{r'} G(\mathbf{r},\mathbf{r'})\left(\beta \nabla' H_0(\mathbf{r'}).\mathbf{m(\mathbf{r'})}-\nabla.\mathbf{m}(\mathbf{r'})\right)\right)
\end{equation}

Which is already a familiar result in the recent literature\cite{moyses_2015,noh_2015}, though we here display it in terms of the Green's function.

\subsection{Conservative and non-conservative perturbations} \label{sec:pp}
In the above, we showed the modification to the Boltzmann distribution from the action of a generic additional vector field of forces $\mathbf{m}$. The modified distribution can be seen to be equal to the Boltzmann distribution with an additional series of perturbations for higher order effects, which are represented as integrals over effective \textit{source terms} with an integral kernel derived from the corresponding equilibrium system (the Green's function).  Thus far we kept the form of the additional forces $\mathbf{m}$ general. We now analyze the nature of this perturbation with respect to two different extremes, one of which is that the additional forces are themselves conservative, and another where the additional forces are divergence free, corresponding to a ``purely" non-conservative perturbation.


Firstly, we imagine that the perturbation is solely made up of a gradient part $\mathbf{m}(\mathbf{r})=-\nabla H_1(\mathbf{r})$. For this special example we can calculate the perturbative terms exactly, for example the first order perturbation given by:
\begin{equation}
\mu_1 = -\mu_0 \beta  \int G(\mathbf{r},\mathbf{r'})\beta^{-1}q_2(\mathbf{r'}) d\mathbf{r'} \label{eq:conpert}
\end{equation}
where for a conservative perturbation the function $q_2$ from equation \ref{eq:q2} is given by:
\begin{equation}
\beta^{-1} q_2=\nabla^2 H_1 - \beta \nabla H_0.\nabla H_1 
\end{equation} 
However the Green's function is just the inverse of this operator acting on $H_1$, thus the integral in expression eq. \ref{eq:conpert} becomes simply  $-\mu_0 \beta H_1$, which is what we would expect from ordinary perturbations to the original Hamiltonian with an extra potential term. In fact it is possible to prove that the full series expression for $\mu$ when the imposed field is conservative is given by:
\begin{equation}
\mu=\exp(-\beta H_1)
\end{equation}
In other words, the Hamiltonians are additive, as we might have expected. When the imposed field is divergence free, $\nabla.\mathbf{m}(\mathbf{r})=0$, with no gradient component, simple closed form expressions cannot be found, however we note the source term has the form $q_2(\mathbf{r}) = \beta^2 \nabla H_0(\mathbf{r}).\mathbf{m}(\mathbf{r})$. From here on we shall refer to this term $\nabla H_0(\mathbf{r}).\mathbf{m}(\mathbf{r})$ simply as ``the source", for if this expression is equal to zero, the entire coordinate space distribution will reduce back to the Boltzmann distribution.

\section{Approximations for probability distributions in non-conservative systems} \label{sec:appr}

We have in the previous section displayed the coordinate space probability of a non-conservative system in terms of a perturbative expansion in the non-conservative forces. However, the form of the distribution is very complicated for the purposes of calculation. As we are not so much interested in exact mathematical results as opposed to leading order physical effects, in this section we show the validity of the perturbative approach for a linear system (where exact results can be obtained) by comparing how the first order theory compares to the exact solution. We then discuss in more depth the coordinate space Green's functions which we introduced in the previous section, and approximate forms of the microscopic probability distribution.

\subsection{Validity of first order perturbation: Linear systems}

We wish to establish that a perturbative approach in non-conservative fields from the previous section gives approximately the correct solutions, and demonstrate the range of validity of the perturbative expansion. We choose the simplest possible system, which is one where all the forces are linear in the particle positions, for example corresponding to a network of harmonic springs, or a particle in an active bath (see S.I  section S1A). In addition, the linear system admits an easier form of numerical solution as \ref{eq:chi} can be recast into an algebraic ricatti equation, which can be solved using numerical techniques (for details, see S.I. section S2). We impose the force field as $\mathbf{F}=\underline{C} \mathbf{r}$ where $\underline{C}$ is a matrix and we search for solutions that satisfy equation \ref{eq:chi} that look like $\phi\sim\mathbf{r}^T\underline{B}\mathbf{r}$, or more precisely, that the probability distribution has the form $P(\mathbf{r})=\exp(-\phi(\mathbf{r}))=\exp(-\beta \frac{1}{2}\mathbf{r}^T.\underline{B}.\mathbf{r})$ where $\underline{B}$ is a matrix.

We will compare the exact numerical exact solution for a linear system of 10 particles in one dimension, with forces given by:  
\begin{equation}
\mathbf{F}(\mathbf{r})=-k \underline{I}.\mathbf{r}+\delta \underline{M_{S}}.\mathbf{r}+\epsilon \underline{M_{AS}}.\mathbf{r}
\end{equation}
Where $I$ is the identity matrix and $M_S$ and $M_{AS}$ are, respectively, a symmetric and anti-symmetric random matrix of couplings uniformly distributed between $1$ and $-1$ and where every diagonal element is zero, and $k$, $\epsilon$ and $\delta$ are parameters that set the strength of the different matrices. For linear systems, identification of conservative and non-conservative forces is very straightforward, symmetric matrices are conservative, and anti-symmetric matrices are non-conservative.

The first order perturbation with these forces would be given by:
\begin{equation}
P(\mathbf{r})=\exp\left(-\frac{1}{2}\beta \mathbf{r}^T.(k\underline{I}+\delta \underline{M_{S}}).\mathbf{r}-\frac{1}{2}\mu_1\right)
\end{equation}
Where $\mu_1$ is the solution to equation \ref{eq:perturb}:
\begin{equation}
\left(\nabla^2-\beta \mathbf{r}^T.(k\underline{I}+\delta \underline{M_{S}}).\nabla\right)\mu_1(\mathbf{r})=\beta^2 \mathbf{r}^T.\left(k \underline{I}+\delta \underline{M_{S}}\right).\left(\epsilon \underline{M_{AS}}\right).\mathbf{r}
\end{equation}
The solution to this equation is given by:
\begin{equation}
\mu_1=\frac{1}{2}\beta\mathbf{r}^T.(k\underline{I}+\delta \underline{M_{S}})^{-1}.\left(\epsilon \underline{M_{AS}}\right).\left(k \underline{I}+\delta \underline{M_{S}}\right).\mathbf{r}
\end{equation}

\begin{figure}[!t]
\begin{center}
\includegraphics[width=90mm]{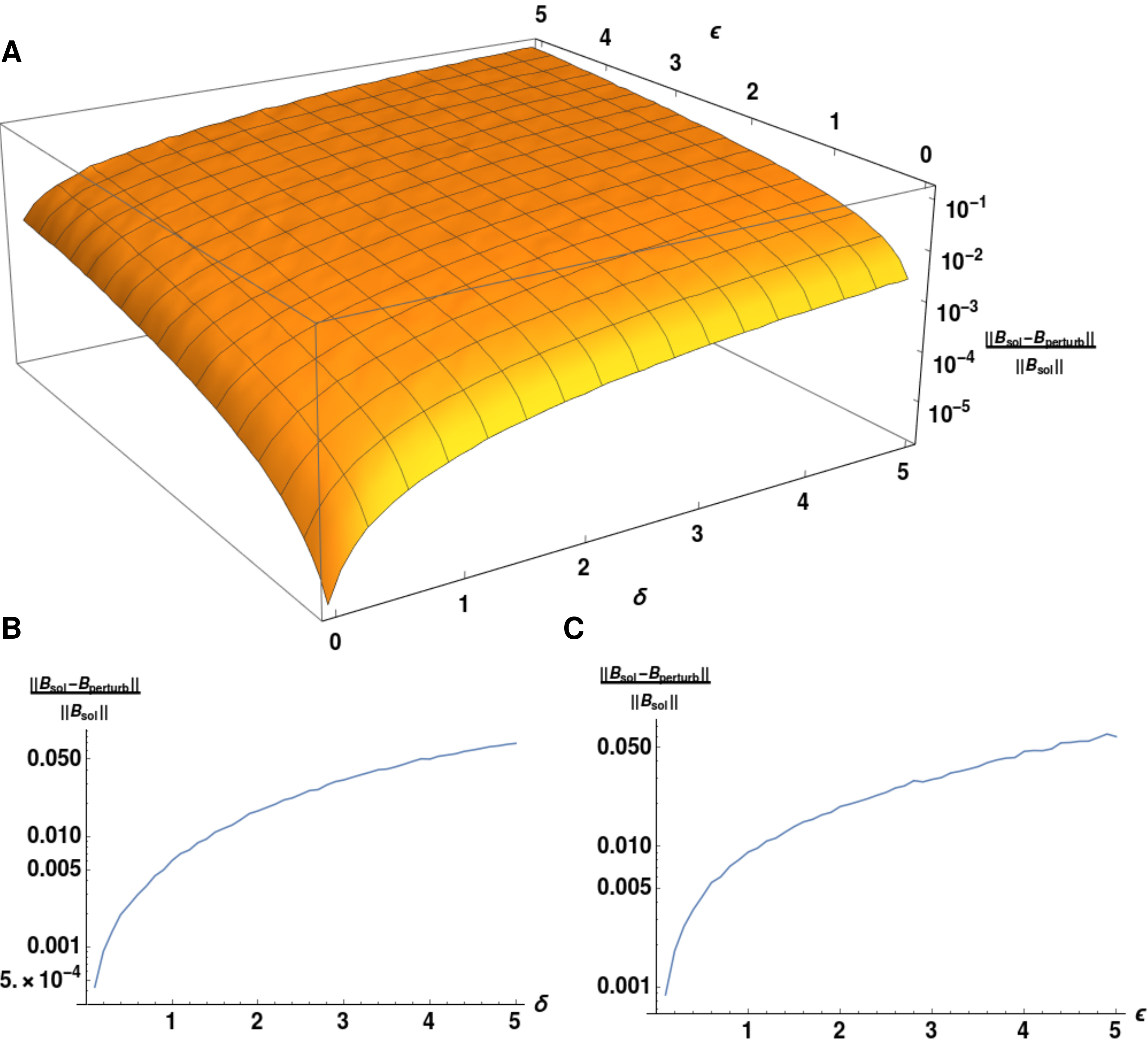}
\caption{Average error between perturbative solutions and real solutions for a linear system for a hundred different realizations of the perturbation matrices $\underline{M}_{S},\underline{M}_{AS}$ for each magnitude of perturbation $\epsilon,\delta$ (with $k=10$), for both first. A) shows the relative error as a function of magnitudes of perturbations $\epsilon$ and $\delta$ are inreased. B) and C) show slices at one point in $\epsilon=2.5$ and $\delta=2.5$ respectively. As expected the perturbation theories become worse for larger perturbations. However, the relative error is acceptable even for large perturbations ($\sim$ 50\% of the magnitude of the equilibrium problem has a $10\%$ relative error in the microscopic probability).}
\label{fig:err} %
\end{center}
\end{figure}

The perturbative and exact solutions to this problem can be written in the form $\frac{1}{2}\mathbf{r}^T \underline{B}.\mathbf{r}$ where $\mathbf{r}$ are the coordinates and $\underline{B}$ is the solution. We compare the similarity of the two solutions as a way of probing the effectiveness of the pertubative approach $||\underline{B}_{\text{sol}}-\underline{B}_{\text{perturb}}||/||\underline{B}_{sol}||$ where $\underline{B}_{\text{sol}}$ is the actual solution and $\underline{B}_{\text{perturb}}$ is the perturbative solution. We calculate the error by taking the matrix norm of the differences divided by the norm of the exact solution. The average error is shown in figure $\ref{fig:err}$ for different magnitudes of the perturbation.

 As seen in figure \ref{fig:err} the error between the real and perturbative solutions for this system becomes larger as the size of the perturbation increases, but it gives good agreement to the real solution even for perturbations which are half the size of the original problem ($\sim10\%$ error).  This is not a rigorous proof that a perturbative expansion always exists in the non-conservative forces, however we also note that for the case where there are only divergence-free non-conservative forces in the system, the probability is constant across all coordinate space (for any force with this property). Therefore, even in the case where the non-conservative forces were very much stronger than the conservative ones, we could imagine performing an expansion around the homogeneous state in order to obtain physically meaningful results.  

\subsection{Coordinate space Green's functions and effective distributions}

We observed in section \ref{sec:analytic} that the full solutions to the system could be represented by integrations of various source terms with Green's function kernels. While a full analytic form of these Green's function kernels is difficult, we can analyze what their effective impact is for realistic systems. The Green's function equation is given by:
\begin{equation} \label{eq:green2}
\left(\nabla^2-\beta \mathbf{\nabla}H_0(\mathbf{r}).\nabla\right)G(\mathbf{r},\mathbf{r'})=\delta(\mathbf{r-r'}) 
\end{equation}
Where in the high temperature limit,$\beta\to0$, the Green's function is given by:
\begin{equation}
\lim_{\beta\to0}G(\mathbf{r},\mathbf{r'})=\frac{c}{\left|\mathbf{r}-\mathbf{r'}\right|^{N-2}}
\end{equation}
with c is a constant, and N is the number of degrees of freedom (for example for a fluid of $n$ particles in three dimensions it is $3n$) . In words, it depends inversely on the distance of points in coordinate space to the power of the total number of degrees of freedom in the system. This is a highly local function. In the series expansion, all the modifications to the microscopic probability are given effectively by:
\begin{equation}
\mu_n(\mathbf{r})\sim\int G(\mathbf{r},\mathbf{r'})\rho_n(\mathbf{r'})\mathrm{d}\mathbf{r'}
\end{equation}

This equation is tricky to solve due to the correlations in the force term $\mathbf{\nabla}H_0$. However, certain simplifications exist. One particularly attractive assumption is that the Hamiltonian is slowly varying in space (see S.I. section S3A). One can then show that the Green's function in these limits is given by:
\begin{equation} \label{eq:simpgreen}
G(\mathbf{r},\mathbf{r'})\approx c\frac{\exp\left[\frac{\beta}{2} H_0\left(\mathbf{r}\right)-\frac{\beta}{2} H_0\left(\mathbf{r'}\right)\right]}{|\mathbf{r}-\mathbf{r'}|^{N-2}}
\end{equation}
This can be shown from equation \ref{eq:green2} with the assumption that higher derivatives of the Hamiltonian are equal to zero. We also mention that as the operator acting on the left hand side of equation \ref{eq:green2} is not self-adjoint for a general Hamiltonian, therefore that the Green's function does not display symmetry in its arguments. Nor does it necessarily form a complete set such that an eigenfunction expansion is always valid.

The form of term in equation \ref{eq:simpgreen} is suggestive of what one may expect for real Green's functions. It has a contribution arising from the term $\nabla^2$, capturing the probabiliy of diffusing from point $r$ to $r'$ which is then weighted by the difference in Boltzmann factors at coordinates $\mathbf{r'}$ and $\mathbf{r}$. Physically, the Green's function characterizes the effect that an impulse at some point $\mathbf{r'}$ has at point $\mathbf{r}$. The asymmetry in the weights is then explicative of the structure of the equilibrium landscape, which we can see by rewriting the weights as $\exp\left(-\beta\left(H(\mathbf{r'})-H(\mathbf{r})\right)\right)$, it is clear that if there is a unit impulse at a point $\mathbf{r'}$ the effect at point $\mathbf{r}$ will be minimal if $H(\mathbf{r})<H(\mathbf{r'})$, i.e., if point $\mathbf{r}$ has lower energy, or is more likely at equilibrium. However, the converse is not true, in that the effect of a unit impulse at a local minimum will strongly affect all the points around it. Moreover, the effect of a unit impulse at another point will also be weighted by the inverse of the distance in coordinate space between them, so the effect will also be very minimal if the points are very far from each other.

Several other consequences of physical significance are apparent from Green's functions. The first of which is that if one wished to calculate the full effect of introduction of a non-conservative perturbation at a point $\mathbf{r'}$ one would need to perform a full integration over all the particle coordinates in order to calculate the new microscopic probability at any point, a result known from discrete systems\cite{maes_2013}. The difference in the effect of a non-conservative perturbation as opposed to a conservative one is that a non-conservative peturbation is non-local in its effect on the probability distribution, as we saw in section \ref{sec:pp} that conservative perturbations lead to a modification of the Boltzmann weight that depends only on the local Hamiltonian (scalar field) at that point.

\begin{figure}[!h]
\begin{center}
\includegraphics[width=90mm]{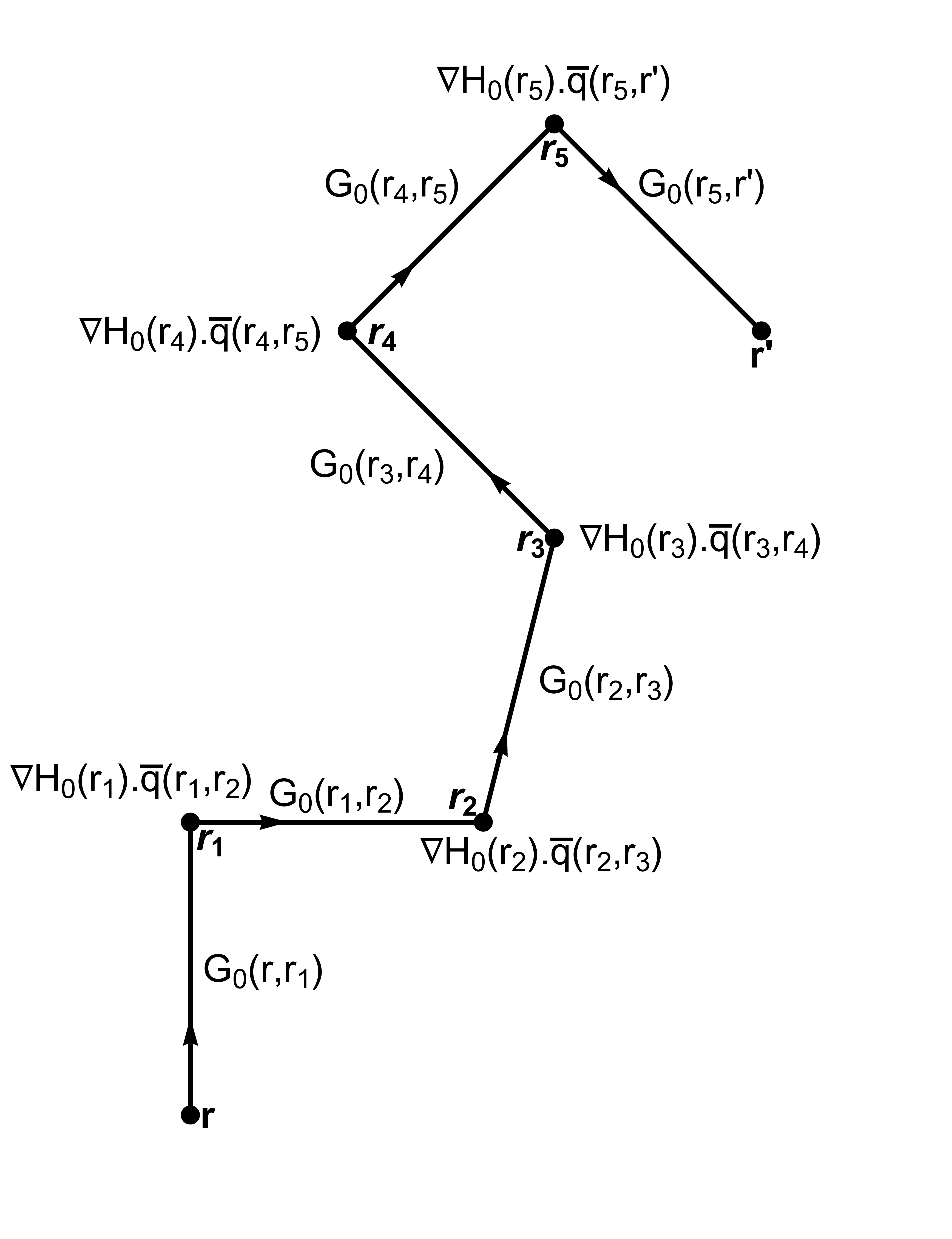}
\caption{Graphical representation of the $nth$ term of the Green's function series solution (in this case $n=6$). One draws a line of $n$ segments connecting the points $\mathbf{r}$ and $\mathbf{r'}$. Associated with each interior vertex is a factor corresponding to the projection of the force at the point along the next line segment. Along each edge is associated a 0th order Green's function. The final form is integrated along all the interior points, giving an effective sum over paths of $n$ links.  }
\label{fig:green} %
\end{center}
\end{figure}

These simple approximations motivate our understanding of the full form of the Green's function. The full Green's function can be written as a Liouville-Neumann series in the Boltzmann factor $\beta$. 
\begin{equation}
G(\mathbf{r},\mathbf{r'})=\sum_{n=0} \beta^n G_n(\mathbf{r},\mathbf{r'})
\end{equation}
The first few terms in this series can be written down as:
\begin{align}
G_0(\mathbf{r},\mathbf{r'})&=\frac{c}{\left|\mathbf{r}-\mathbf{r'}\right|^{N-2}}\\
G_1(\mathbf{r},\mathbf{r'})&=\int \mathrm{d}\mathbf{r_1} G_0(\mathbf{r},\mathbf{r_1})G_0(\mathbf{r_1},\mathbf{r'})\left(-{\nabla}H_0(\mathbf{r_1}).\mathbf{\bar{\mathbf{q}}}(\mathbf{r_1},\mathbf{r'})\right)\\
G_2(\mathbf{r},\mathbf{r'})&=\int \mathrm{d}\mathbf{r_1}\mathrm{d}\mathbf{r_2} G_0(\mathbf{r},\mathbf{r_1})G_0(\mathbf{r_1},\mathbf{r_2})G_0(\mathbf{r_2},\mathbf{r'})\left(-{\nabla}H_0(\mathbf{r_1}).\mathbf{\bar{\mathbf{q}}}(\mathbf{r_1},\mathbf{r_2})\right)\left(-{\nabla}H_0(\mathbf{r_2}).\mathbf{\bar{q}}(\mathbf{r_2},\mathbf{r'})\right)
\end{align}
Where we introduce the vector $\mathbf{\bar{\mathbf{q}}}(\mathbf{r_1},\mathbf{r'})=\frac{\mathbf{r_1}-\mathbf{r'}}{|\mathbf{r_1}-\mathbf{r'}|^2}$. It is perhaps easier to see how each term in this expansion can be written down graphically, which we include in figure \ref{fig:green}. The graphical representation of each term in figure \ref{fig:green} makes more clear the physical aspects of the Green's function. Despite the fact we have not introduced any path integrals in our analysis so far, the solution to the deceptively simple equation \ref{eq:green2} takes the form of integrating over all paths connecting the points $\mathbf{r}$ and $\mathbf{r'}$. This path is weighted by the $0th$ order Green's function, which strongly suppresses overly long paths connecting the two points, and by the factor $\left(-{\nabla}H_0(\mathbf{r_1}).\mathbf{\bar{\mathbf{q}}}(\mathbf{r_1},\mathbf{r_2})\right)$ which weights how ``helpful" the equilibrium forces at each point along the path are, by calculating their magnitude of the projection of the vector to the next point along the path. This full form of the Green's function then shows the impact of a source at a point $\mathbf{r'}$ affects the probability of $\mathbf{r}$, a similar picture emerges as in the slowly varying field approximation, however in this equation the path dependence of the Green's function becomes clear, in that the effect of a source at point $\mathbf{r'}$ is strong if there is a path of ``helpful" forces connecting the two points. The slowly varying approximation should then be seen to replace the complicated sum over paths of force projections to the effective magnitude of the Hamiltonian at the end points.

This path integral, which is integrating terms going as $\sim G_0 \nabla H_0$ along all paths would apply in cases even where the equilibrium force is strong. As generally a vector field that goes as $\sim G_0 \nabla H_0$ is not necessarily conservative, we cannot just replace the integrals along the paths by the values at the end points, if this were to be true, we can write down a form of the Green's function which depends only on the endpoints. Curiously, even though equilibrium systems have no path dependence, the Green's function that characterizes modification due to non-conservative perturbations contains information about the sum of all paths at equilibrium. Despite all this complexity, as we mentioned, if the source term is of a particular form, corresponding to a conservative perturbation, we get the local result (section \ref{sec:pp}).

While the full form of microscopic probability would be difficult to calculate, it's net effect for systems which do not have equilibrium forces that are very strong would result in an effective ``smearing" out of whatever the source term is by the Green's function kernel, which would serve to smooth out features over a particular length scale. In addition, the Green's function is highly local, it will be peaked strongly around $\mathbf{r}=\mathbf{r'}$. This naturally leads us to the \textit{locality approximation} applied even to non-conservative source terms, the consequences of which we shall examine in much more detail in the next section, that the operation of integration with the Green's function kernel is effectively local, that is to say the relevant terms after integration with the Green's function kernel will be go as:
\begin{equation}
\int \mathrm{d}\mathbf{r'} G(\mathbf{r},\mathbf{r'})\rho_c(\mathbf{r'})\sim\lambda_1^2 \rho_c(l\mathbf{r})
\end{equation}
Where we introduce extra effective length scale $\lambda_1$ and some scale factor $l$. Note that as these terms are modifying probabilities, any addition or subtraction to $\rho_c$ will not affect the probability distribution after normalization (similar to changing the gauge of a potential, it is only relative differences that are important). In the S.I, we justify this approximation (S.I. section S3B) and we also demonstrate this effective smearing numerically for a simple example (S.I. section S3C). 

Therefore, we claim that to first order an approximate expression for the steady state distribution in the presence of both conservative force $-\nabla H_0(\mathbf{r})$ and non-conservative force $\mathbf{m}(\mathbf{r}),\text{with} \nabla.\mathbf{m}(\mathbf{r})=0$ is given by:
\begin{equation} \label{eq:noneq}
P(\mathbf{r})\approx\exp\left(-\beta H_0(\mathbf{r})+\beta^2 \lambda_1^2 \nabla H_0(\mathbf{r}).\mathbf{m}(\mathbf{r})\right)
\end{equation}
This expression is also found to also be a good approximation to the linear system under the same conditions under which the perturbative approach is valid. We expect it to be qualitatively valid in systems which are not too strongly correlated at equilibrium, and where the non-conservative effects are perturbative.

\section{Larger scale effects: non-conservative systems and shape} \label{sec:shape}

The preceding sections were rather abstract in their application, of more immediate importance in experimental realizations would be the larger scale properties of such systems. In this section we will try to more phenomenologically understand how \textit{microscopic} driving will lead to \textit{macroscopic} effects. The correct macroscopic degrees of freedom for non-equilibrium system are not obvious\cite{solon_2015}, so we will choose to focus on the density of the particles, which always exists and is well defined in and out of equilibrium. Rather than treating the density as a conserved field and reasoning over the form of the density currents from physics in a continuity equation like scheme, we instead wish to focus on how our microstate probability \ref{eq:noneq} would modify the macroscopic density that one would observe for any non-conservative system.

We introduce by hand such systems interacting with both conservative and non-conservative parts by introducing an extra pair force between particles which is given by a curl of some vector $\mathbf{m_{ij}}(\mathbf{r}_{ij})=\nabla_i\times \mathbf{A}(\mathbf{r}_{ij})$, but stress these are not meant to represent real physical systems. In particular by choosing a pair non-conservative force arising in this way we have chosen axes of the system. The closest physical picture that would correspond to this kind of system would be one in which a global field (such as a magnetic field) induces a constant angular momentum in the particles relative to the imposed field, and then these angular momenta interact with each other according to some effective curl, reminiscint of some experimentally realized systems\cite{soni_2019}. In order to model real active matter systems, we would have to introduce particles with their own orientational degrees of freedom, Alternatively, non-conservative three body forces between particles could be introduced. This notwithstanding, these models can show us qualitatively the effect that non-conservative microscopic forces have on spatial distributions of the system.  We are particularly interested in whether any generalities arise from the consideration of the modified Boltzmann distribution in equation \ref{eq:noneq}.

As we mentioned, the observable quantity we are most interested in is the density of a fluid of particles interacting non-conservatively, which corresponds to the statistical average of the density operator under the stationary distribution:
\begin{align}
\rho(\mathbf{x},\mathbf{r})&=\sum_{i=1}^N \delta(\mathbf{x}-\mathbf{r_i}) \\
\langle\rho(\mathbf{x})\rangle &=\int \mathrm{d}\mathbf{r} \rho(\mathbf{x},\mathbf{r})P(\mathbf{r});
\end{align}
where $\mathbf{x}$ now refers to ordinary three dimensional space. In particular we are interested in how densities differ for systems with non-conservative microscopic forces. In equilibrium there is a well defined connection between the free energy of the system and this density, which is captured by a free energy functional\cite{evans_1979}:
\begin{equation}
F\left[\rho(\mathbf{x})\right]=\beta^{-1}\int\mathrm{d}\mathbf{x} \rho(\mathbf{x})\left(\log(\lambda_{TH}^3 \rho(\mathbf{x}))-1\right))+\frac{1}{2}\iint \mathrm{d}\mathbf{x}\mathrm{d}\mathbf{x'} \rho(\mathbf{x})\rho(\mathbf{x'})U(\mathbf{x}-\mathbf{x'})
\end{equation}
Where $U$ is some pair potential acting between the particles. The second term of this equation, arising from interparticle potentials, can be derived from reformulating the Hamiltonian $H=\sum_{i\ne j}^N U(\mathbf{r_i}-\mathbf{r_j})$ using the density operator. The density field which minimizes this functional is the best approximation to the true density of a system of particles interacting with the potential $U$.

We can reason about what the effect of pairwise non-conservative driving is by considering the effective source term that appears in the modified microscopic probability $\nabla H_0.\mathbf{m}$. There are several salient facts about this source term, the most important of which is that if it is equal to zero, either from the non-conservative force being zero, or the angle between the non-conservative force and the conservative force is always equal to zero, then the stationary properties of the system will be given by a Boltzmann distribution arising from the conservative part. This source can also be described using the density operator in the same way as the equilibrium Hamiltonian, using pairwise addititive curl force between particles (ignoring density fluctuations):
\begin{equation}
\nabla H_0.\mathbf{m}\approx\iiint \mathrm{d}\mathbf{x}\mathrm{d}\mathbf{x'}\mathrm{d}\mathbf{x''} \rho(\mathbf{x})\rho(\mathbf{x'})\rho(\mathbf{x''})\left(\nabla U(\mathbf{x}-\mathbf{x'}).\nabla\times A(\mathbf{x}-\mathbf{x''})\right)
\end{equation}
Where now the differential operators $\nabla$ are with respect to ordinary 3 dimensional space.

In the previous section, we saw that the first order approximation for the microscopic probability included an additional term in the exponent of the Boltzmann distribution of the form $\nabla H_0.\mathbf{m}$. We introduce a qualitatively the effect of this driving by introducing the following functional:
\begin{align}
F^{\text{(nc)}}=&\beta^{-1}\int\mathrm{d}\mathbf{x} \rho(x)\left(\log(\lambda_{TH}^3\rho(\mathbf{x}))-1\right))+\frac{1}{2}\iint \mathrm{d}\mathbf{x'}\mathrm{d}\mathbf{x} \rho(\mathbf{x})\rho(\mathbf{x'})U(\mathbf{x}-\mathbf{x'})\\&+\lambda_1^2 \beta \iiint \mathrm{d}\mathbf{x''}\mathrm{d}\mathbf{x'}\mathrm{d}\mathbf{x} \rho(\mathbf{x})\rho(\mathbf{x'})\rho(\mathbf{x''})\nabla U(\mathbf{x}-\mathbf{x'}).\nabla\times\mathbf{A}\left(\mathbf{x}-\mathbf{x''}\right)
\end{align}
While we label this $F$, we wish to be careful in calling this a free energy, as this system is inherently not in equilibrium. However we assume a similar mapping exists from a qualitative description of the microscopic probability in our non-conservative case to an effective functional of the particle density (with all the subtleties this entails). The first term of this functional is the familiar ideal gas contribution. The next encompasses the (conservative) pairwise energy between particles in our system. The next corresponds to modifications due to the non-conservative forces, which exist as an effective three body potential\cite{iyetomi_1989} of special form.

While the contribution to the free energy due to interparticle potentials is fairly unconstrained, the term corresponding to non-conservative pair forces is more interesting. A salient mathematical fact of integrals of this type is that:
\begin{equation} \label{eq:zerocond}
\int_{\text{all space}} \mathrm{d}\mathbf{x} \nabla U(\mathbf{x}).\nabla\times\mathbf{A}(\mathbf{x})=0
\end{equation}
Therefore the contribution to this functional arising from \textit{any} non-conservative forces will always be equal to 0 when the density is constant. Taking this logic further, pairwise non-conservative interactions, if they do anything to the density, would be to drive the density profiles of the system from the homogeneuous state. For real physical systems being internally driven, we would have to consider higher order terms in order to capture the relevant physics, however a similar picture would emerge for terms which depend on this source squared and so on. This interesting result arises solely from the fact that the forces are non-conservative, and decomposable in terms of curls. 

\begin{figure}[!b]
\begin{center}
\includegraphics[width=90mm]{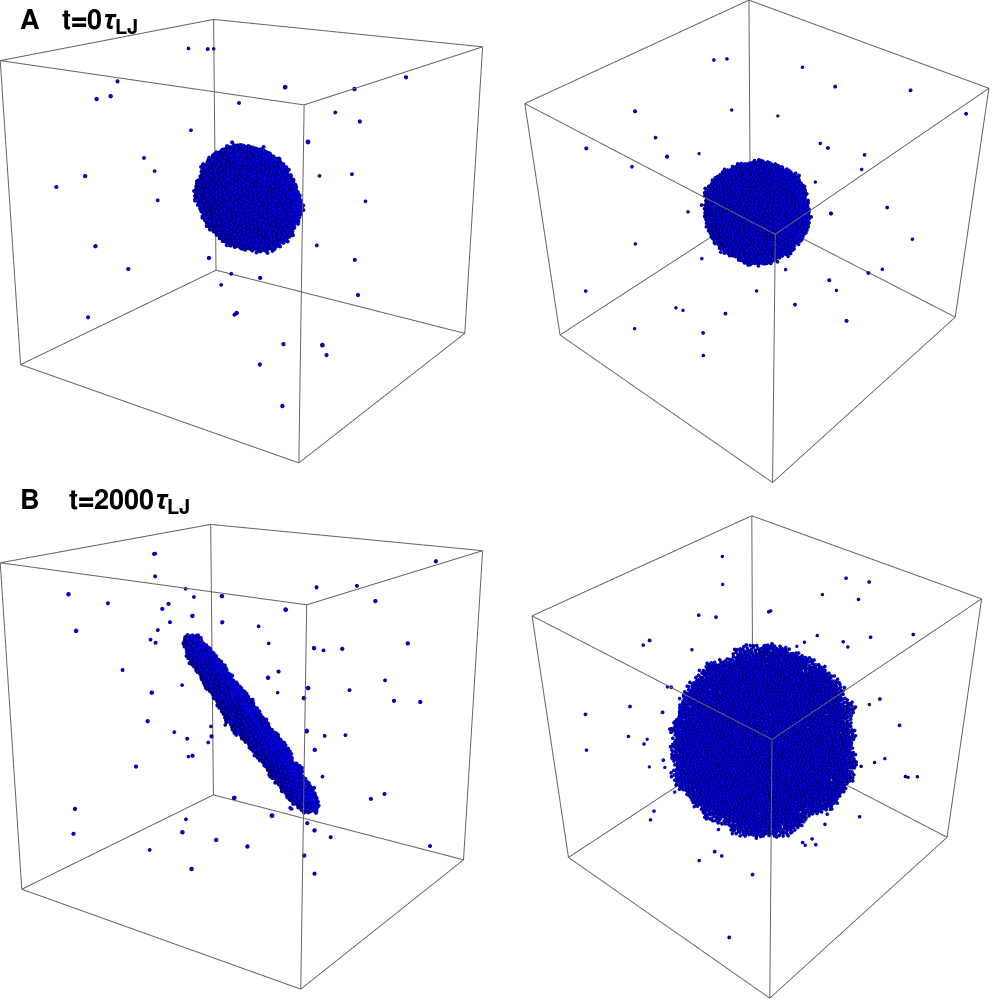}
\caption{Snapshots of simulations of particles interacting under a Lennard-Jones potential and a pair non-conservative force given in equation \ref{eq:pairnc} . In A) we show different viewpoints of the startpoint of the simulation with a roughly spherical droplet of the particles, after relaxation under only the Lennard-Jones potential and immediately after imposition of the non-conservative force. In B) we show the stationary droplet shape after time evolution of the microscopic equations after switching on the non-conservative forces. The particles, which would under the ordinary (equilibrium) Lennard-Jones form a spherical droplet, are instead driven away from the spherical shape through their non-conservative interactions. Adopting an elliptical shape with respect to the axes where their internal degrees of freedom are fixed.  }
\label{fig:ncs} %
\end{center}
\end{figure}

We can see how this effective functional changes for particular kinds of density field $\rho(\mathbf{x})$. If the source has a finite length scale and the density is approximately constant over this length scale, then the contributions to $F^({nc})$ from these terms are effectively zero (ignoring fluctuations). Therefore, the main contribution to $F^{(nc)}$ from non-conservative forces happens due to \textit{surfaces}. This can be proven using the divergence theorem for a constant but finite density field, however we can demonstrate this using simulations with pairwise non-conservative forces. In order to test this we simulate 10000 particles in a box with periodic boundary conditions interacting with both an ordinary Lennard-Jones potential:
\begin{equation}
U(r)=4\epsilon_{LJ}\left( \left(\frac{\sigma}{r}\right)^{12}-\left(\frac{\sigma}{r}\right)^{6}\right)
\end{equation}
and where we now introduce an extra divergence-free force between the particles, again not meant to represent a real physical system, but rather as a qualitative representation for how non-conservative forces affect spatial density:
\begin{equation} \label{eq:pairnc}
\mathbf{F}_{ij}(x_{ij},y_{ij},z_{ij})=\frac{\epsilon}{\lambda^2}\nabla_i\times\frac{1}{\lambda^2}\left(\left(x_{ij},y_{ij},z_{ij}\right).f(s_i,s_j).\left(x_{ij},y_{ij},z_{ij}\right) \exp(-(x_{ij}^2+y_{ij}^2+z_{ij}^2)/\lambda^2)\right)
\end{equation}
Where $f(s_i,s_j)$ would correspond to a matrix that depends on some microscopic properties of the particles (such as orientation $s$). Instead of treating these internal degrees of freedom fully in a simulation, we fix the form of the matrix with respect to the system coordinates, somewhat like imposing a strong field aligning the particles. In order to show qualitative variation from expected equilibrium density profiles, we first simulate the relaxation of the system to an equilibrium state, and then switch on the non-conservative forces. In addition, as we expect non-conservative forces to mainly affect surfaces, we keep the total particle volume fraction of the system small $\frac{N \pi \sigma^3}{6 V}=0.01$. The other parameters we choose are that $k_b T=1$, $\lambda=1.5$, $\sigma=1$,$\epsilon_{LJ}=2.0,\epsilon=1$

We show a particular example of the shape adopted by this system in figure \ref{fig:ncs}. This simple simulation already shows qualitatively the effects that we were able to reason about through a consideration of the source term. The extra term in the effective ``free energy" drives the system away from its equilibrium shape (which would be a spherical droplet), leading to the formation of a pancake like shape. It should be noted that while this appears to be similar to elastic deformation of a sphere, the system is heavily overdamped and the states arise from microscopic pair forces. From the preceding discussion we stated that the source term $\nabla H_0.m$ would act primarily through surfaces. We can now quantify this effect by looking at the same snapshots of system evolution with the color scale of the particles equal to their total value of this source and also see how the average value of this term changes during our simulation run. 

\begin{figure}[!h]
\begin{center}
\includegraphics[width=140mm]{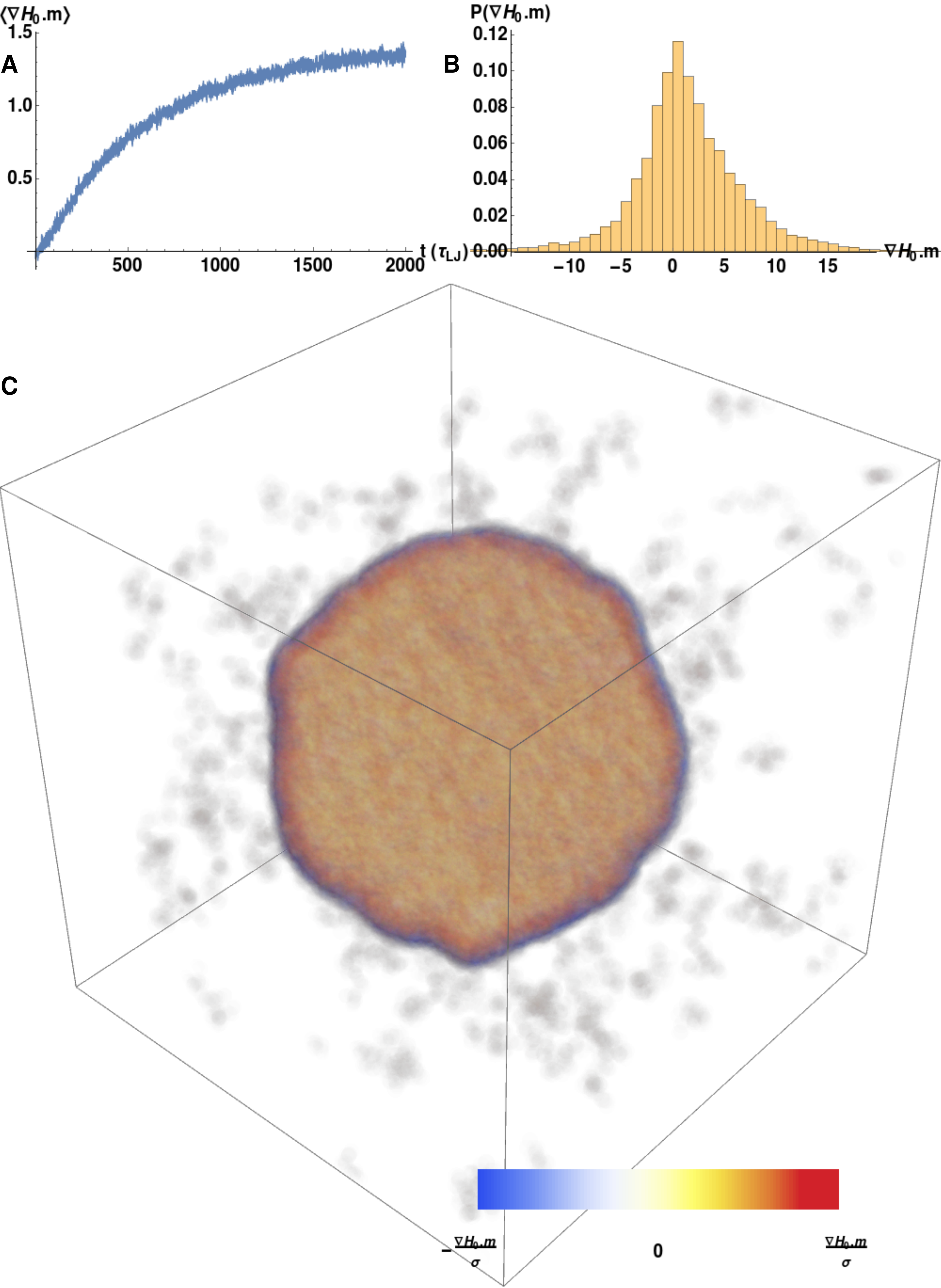}
\caption{Characterization of the final states of the system in terms of the effective sources we defined in the main text. (A) The average value of the source term $\nabla H_0.\mathbf{m}$ during the run of the simulation snapshots seen in figure \ref{fig:ncs}. The total value of the source term increases with simualtion time, and saturates at the pancake like shape shown in the figure \ref{fig:ncs}. (B) The probability distribution of the source term, it is skewed somewhat towards positive values. (C) How the value of the source is distributed over space, with a color scheme selected from the standard deviation of the distribution. One can observe that the far outer edge of the pancake is blue, corresponding to negative values of the sourc term, but all the surfaces inside are slightly positive. There exists a sub surface layer that is also positive, below the outer edge. A similar figure produced for the initial state in figure \ref{fig:ncs} would show a source term which is approximately zero for all the particles.  }
\label{fig:source} %
\end{center}
\end{figure}

These results are presented in figure \ref{fig:source}.  Several trends are visible from this figure. Firstly, that the mean value of $\nabla H_0.\mathbf{m}$ per particle changes during the simulation as it moves from the spherical shape to the pancake shape. However, the distribution of $\nabla H_0.\mathbf{m}$ over all the particles is very broad compared to the mean. Moreover, these values are not distributed equally across the final state, but are instead localized in different parts of the pancake. As can be seen, there exists a strong surface layer across one axes of the pancake, and a differently valued one across another surface, according well with the mathematical arguments which we presented earlier. 

This simple phenomenology and simulations show us qualitatively the effect non-conservative microscopic interactions can have on larger scale observations. We kept our discussion general up until the point of simulating the particles. Interestingly, a natural result of a framework considering the interplay of microscopic non-conservative interactions was that it will subsequently manifest itself through density inhomogeneities, and most strongly through surfaces. In particular, the imposition of given forms of microscopic driving naturally lead to particular shapes when considering aggregates of such particles, given by the interplay of the conservative and non-conservative forces. In this system the non-conservative and conservative forces compete against each other over which shape they find preferable, settling on the pancake shown.

 \clearpage
\clearpage

\section{Discussion} \label{sec:diss}

We have in the preceding displayed results relating to both the microscopic and macroscopic treatments of non-conservative forces in a system of particles. We will now proceed to discuss some broader aspects of the physical relevance of these results, in both its microscopic and macroscopic manifestations.

\subsection{Physical intuition of microscopic perturbations to Boltzmann distributions by non-conservative vector fields}

\begin{figure}[!h]
\begin{center}
\includegraphics[width=140mm]{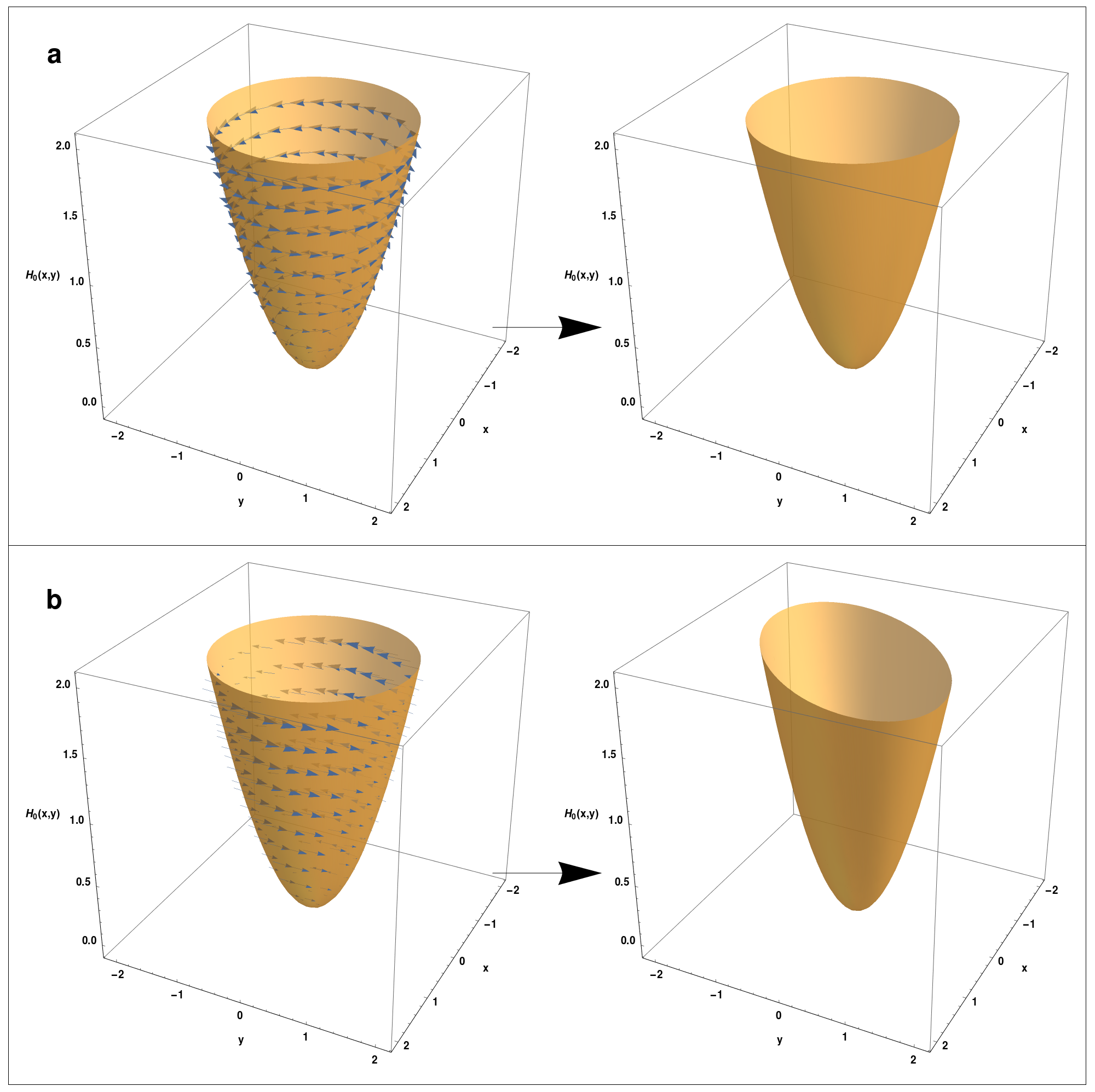}
\caption{Impact of imposing a non-conservative field on an harmonic equilibrium system $(-\epsilon y,x)$ (a) for $\epsilon=1$ the probability distribution is unmodified as the non-conservative field and the equilibrium forces always move orthogonal to each other (the divergence of the non-conservative field is zero) (b) as $\epsilon$ is varied, the vector fields are no longer orthogonal, so the distribution is modified. The non-conservative vector field ``pushes" against the equilibrium one, leading to modifications of the probability distribution. }
\label{fig:dis} %
\end{center}
\end{figure}

The linear system provides a useful avenue to test our intuition regarding the effects of non-conservative force. We imagine the coordinate space of the equilibrium problem as a surface embedded in 3 dimensions, where the $x,y$ axes corresponds to different points in coordinate space and the height in $z$ is the energy. For a harmonic system we plot the coordinate space of the system and imagine what would happen upon imposition of a non-conservative vector field at every point on the surface. It is easy to see visually (figure \ref{fig:dis}) that the average effect of this vector field will be zero if the divergence is zero and the dot product of the non-conservative field with the equilibrium forces is zero. For such a non-conservative vector field, the motion due to that force in coordinate space proceeds along the equilibrium constant energy surface, and the probability along each point on that surface is not modified.

This helps to motivate our understanding of the ``source" term, $\nabla H_0.\mathbf{m}$, which we refer to throughout the text. In a system being driven out of equilibrium out by non-conservative forces, states where the non-conservative force is balancing the equilibrium one are stabilized with respect to the equilibrium distribution. This is by itself not a particularly complex idea, though its manifestations can lead to rather complicated effects. Every term in the perturbative expansion depends on this source, so if it is zero the system will relax to an equilibrium distribution over states.

We also note that while this term would appear in the total dissipated power of the system, given by $\left\langle \sum_i F_i \frac{\mathrm{d} x_i}{\mathrm{d} t} \right\rangle$, it is not by itself the total dissipated power of the system\cite{tom_2006}. We have in this paper framed our approach towards non-equilibrium steady states in terms of analyzing the dynamics of systems with unspecified non-conservative forces. Another approach, which we did not take in this manuscript, would be to analyze systems  such more generically in terms of quantities such as entropy production, work and dissipation. Some of the same concepts seem to arise from the microscopic picture, as should be expected, however, future work should relate both the thermodynamic picture and the microscopic picture presented here to each other.

\subsection{Microscopic non-conservative forces and larger scale effects}

Proceeding from microscopic consideration of non-conservative forces led to particular features when analyzing average properties such as the density. Our simple phenomenological picture led to the coupling of terms in the effective non-equilibrium ``free energy" to surfaces of the density. However, in this simple picture we also ignored the effects of density fluctuations, which non-conservative forces would also affect. 

Despite this, even the simplest phenomenological consideration of non-conservative forces results in the system having a preference for selecting certain density profiles, as was seen in the results of section \ref{sec:shape}. A natural consequence of non-conservative over conservative perturbations is that non-conservative perturbations couple to surfaces (inhomogeneities) rather than to bulks. This is the main difference to a conservative perturbation, which would modify effective free energies in the bulk phase as well.

In the examples we have chosen, the conservative (equilibrium) interactions would lead to the formation of a spherical droplet. The introduced pairwise non-conservative interactions, which favor surfaces, leads to the formation of a pancake phase. The interplay of bulk favoring conservative interactions and surface favoring non-conservative interactions could in principle lead to many unique and interesting steady states. We have here shown only one example, but the principle perhaps could be extended further to understand interactions among multiple species. Moreover, the way in which complex equilibrium interactions supplemented with microscopic driving between components could manifest themselves macroscopically could be suggestive for analysis of biological systems. In principle, this defines a route to attempt to understand the formation of complex non-equilibrium patterns from a consideration of microscopic forces. Intriguingly, the fact that non-conservatives forces couple strongly to density surfaces suggests that complicated systems, with many species, where there are a lot of effective ``surfaces" may behave in ways that are very different to single component fluids which are also being driven out of equilibrium microscopically. 

At this juncture we will claim these ideas as more suggestive than complete, a full theory of the translation of the microscopic driving forces to macroscopic effects would need several ingredients to be considered formal. Firstly, the green's functions locality approximations that have been made would have to be shown more quantitatively to be robust, or if they are system specific it would be necessary to characterize exactly the points at which they fail. Secondly, a full incorporation of the way which non-conservative forces couple to density fluctuations would be required.

\subsection{Active matter and force decomposition}

In this paper, instead of solving for a particular active system, we have presented results in terms of conservative and non-conservative force fields. This approach is not generally standard in current treatments of active matter. However, we argue that such analysis is not without merit. Particular models of active matter may display the features we have discussed here. We created non-conservative systems by construction
in section \ref{sec:shape}. For a real system, we would instead proceed by writing down the realistic forces, and then we could attempt to decompose the forces mathematically into their conservative and non-conservative parts. While this may sound complicated, the fact that in most realistic systems forces are pairwise between particles would reduce the complexity of the problem massively and leave it amenable to numerical fitting techniques if mathematical solutions are impossible. We would expect such systems to qualitatively obey the distributions we described here. However, an extension of the currently presented theory would have to be derived in order to describe systems coupled to different temperature baths. In this way we treat the problem of finding the steady state distributions of active matter systems as the statistical mechanics of non-conservative vector fields, and consider equilibrium as the statistical mechanics of conservative vector fields.

In many other active systems, the non-conservative forces exist in degrees of freedom that aren't necessarily spatial, for example internal degrees of freedom such as orientation may apply effective forces in space. The ideas presented here would have to reformulated for such systems, though effective surface terms corresponding to some more complicated interplay of spatial and internal degrees of freedom may exist.

\section{Conclusion and Future Directions}

In this paper, we have analyzed the steady state distribution that arise from particles interacting with both conservative and non-conservative forces. We have shown that the Boltzmann distribution is modified by the addition of excess convolution integrals over effective source terms that arise when treating the system pertubatively in the non-conservative elements. Finally, we give some indication for how non-conservative forces affect more macroscopic observables, and find that generic non-conservative forces couple to density surfaces rather than volume terms. The consequences of this are that microscopic driving implemented between pairs of particles leads to the adoption of different density profiles and shapes than would be expected from the same system with only the conservative interactions. However, the fact that such deformations arise can be understood from the source terms we derived in the microscopic theory, where differences from Boltzmann distributions arise due to operators acting on $\nabla H_0\left(\mathbf{r}\right).m\left(\mathbf{r}\right)$, and this perturbation, when course grained, naturally couples to surfaces. We believe the framework we have presented so far is suggestive of future directions.

On the mathematical level, the full microscopic probability distribution is very complicated. The probability distribution due to non-conservative forces can be written as a series of integrals over source terms with Green's functions, where the Green's function is some equilibrium path integral. Ignoring this path integral and treating the effects of non-conservative forces as a local modification of probability was seen to approximately reproduce real probabilities, and was also qualitatively capture features on the larger scale.


While in this manuscript we have mainly focused on the static distributions of matter being driven by non-conservative forces, a whole other aspect that was ignored in this work was the dynamics. The steady states arising from these forms of non-equilibrium systems may still have net currents acting. How these steady states then behave when perturbed by an external field, their response properties and dynamic relaxation are all ongoing topics of research\cite{maes_2008}. An additional assumption we employed was that the system was overdamped. Upon relaxation of this assumption many other interesting phenomena may result.

In the latter sections, we noted that non-conservative systems adopt particular shapes compared to conservative ones. We believe this is an interesting avenue of potential exploration. In particular, one may be able to relate particular mesoscopic patterns or shapes to the interplay of microscopic conservative and non-conservative forces. Additionally, one may be able to reformulate this theory in terms of conservative and non-conservative stress tensors \cite{krger_2018}.

\section{Acknowledgements}

The author would like to the thank Hridesh Kedia, Peter Foster and Yitzhak Rabin for helpful discussions. This work was funded through the Gordon and Betty Moore foundation.

\bibliographystyle{unsrt}
\bibliography{exported-references}

\end{document}


\title{Supplementary information for spatial distributions of non-conservatively interacting particles}
\author{
Dino Osmanovi{\'{c}}
}
\affiliation{Center for the Physics of Living Systems, Department of Physics, Massachusetts Institute of Technology, Cambridge, MA 02139, USA}%
\date{\today}
\maketitle

\section{Non-conservative active matter}
In this section we shall illustrate how two different classes of active system can be understood as systems evolving under a non-conservative force.

\subsection{Colloid in a bacterial bath}
For one of the simplest examples of an active system, for example an optically trapped colloidal bead inside a bath of bacteria, whose time evolution of position $x$ can be modelled through the following equation:
\begin{equation}
\gamma \frac{\mathrm{d} x(t)}{\mathrm{d} t}=-k x(t)+\xi(t)+A(t)
\end{equation}
where we assume that this colloid is overdamped and exists in a media with a dissipative term $\gamma$ and a white noise term $\xi$ which has the property that $\langle \epsilon(t)\epsilon(t')\rangle =2 k_b T \gamma \delta(t-t')$. Supplementing this is an active noise $A(t)$ where $\langle A(t)A(0) \rangle=\frac{C}{\tau_a}\exp(-t/\tau_a)$ where $\tau_a$ is the timescale of some active noise. By representing the active forces in terms of an extra degree of freedom we get an expression in two degrees of freedom describing the same process:
\begin{align}
\gamma \frac{\mathrm{d} x(t)}{\mathrm{d} t}&=-k x(t)+\xi(t)+A(t)\\
\frac{\mathrm{d} A(t)}{\mathrm{d} t}&=-\frac{1}{\tau_a}A(t)+\xi_2(t)
\end{align}
where we introduce another fluctuating white noise process $\xi_2(t)$, which will have the same correlation property as $\xi$ but possibly with a different magnitude. This system can be defined in terms of a matrix of forcings:
\begin{equation}
\frac{\mathrm{d}}{\mathrm{d} t}\left(
\begin{array}{c}
 x(t) \\
 A(t) \\
\end{array}
\right) = \left(
\begin{array}{cc}
 k & 1 \\
 0 & \frac{1}{\text{$\tau $a}} \\
\end{array}
\right)\left(
\begin{array}{c}
 x(t) \\
 A(t) \\
\end{array}
\right)+\left(
\begin{array}{c}
 \xi(t) \\
 \xi_2(t) \\
\end{array}
\right)
\end{equation}

Where the deterministic forces are non-conservative in the space $(x,A)$, defining the system as an out of equilibrium one in these expanded state space.

\subsection{Chemically active systems}

In this section we look at the conditions under which 3 body forces with modifiable interactions, which is a simple model of a chemically active system\cite{osmanovi_2019}, are not conservative. The forces in this system are given by:
\begin{equation}
\mathbf{f}=-\nabla H_0+\mathbf{m}
\end{equation}
Where the vector $\mathbf{m}$ gives us the modifiable forces. Each element of this vector is given by:
\begin{equation}
m_k=\sum_{i \neq k}M_{ki}(\epsilon_{ki},\mathbf{r})\mathbf{F}_{ki}^{(m)}
\end{equation}
in other words, we have a sum of forces acting on each particle, but the magnitude of these forces depends on the state through the functions $M$. By working with the effective Hamiltonian:
\begin{equation}
H_m=\sum_{i \neq j} M_{ij}(\epsilon_{ij},\mathbf{r})\phi^{(m)}(r_{ij})
\end{equation}
where the pairwise force in the vector $\mathbf{m}$ is given by the gradient of some potential $\mathbf{F}_{ki}^{(m)}=-\nabla \phi^{(m)}(r_{ij})$. In terms of this effective Hamiltonian, the total forces on the system can be written as:
\begin{equation}
\mathbf{m}=-\nabla H_m + \mathbf{c}
\end{equation}

where now we have added a vector $c$ which is given by:
\begin{equation}
\mathbf{c}=\sum_{i>j} \nabla(M_{ij}(\epsilon_{ij},\mathbf{r}))\phi^{m}(r_{ij})
\end{equation}

Therefore, by looking at whether the impact of chemical modification, through the vector $c$, is conservative or not, can tell us a lot about the subsequent behaviour of the system. In order to illustrate this we will look at a system with 3 particles in one dimension because the condition to check whether the vector field $\mathbf{c}$ is conservative is most easy to analyse in three dimensions using the condition that the curl of $\mathbf{c}$ is zero, $\nabla \times \mathbf{c}=0$ means the system is conservative. We implement the chemical activity functions as follows:
\begin{align}
M_{12}(\epsilon_{12},\mathbf{r})&=f_1(\mathbf{x}) \\
M_{23}(\epsilon_{23},\mathbf{r})&=f_2(\mathbf{x}) \\
M_{13}(\epsilon_{13},\mathbf{r})&=f_3(\mathbf{x})
\end{align}
Where the vector $\mathbf{x}$ is the vector of all the particle positions $(x_1,x_2,x_3)$, and $x_{ij}$ is used to denote the distance between particles $i$ and $j$.

Using these to calculate $\mathbf{c}$, with a generic pair potential between all the particles $\phi(x_{ij})$, implementation of the curl gives the following condition for conservativeness:
\begin{equation}
\frac{\partial \phi(x_{12})}{\partial x_2}\frac{\partial f_1(\mathbf{x})}{\partial {x_3}}+\frac{\partial \phi(x_{23})}{\partial x_2}\frac{\partial f_2(\mathbf{x})}{\partial {x_3}}-\frac{\partial \phi(x_{23})}{\partial x_3}\frac{\partial f_2(\mathbf{x})}{\partial {x_2}}-\frac{\partial \phi(x_{13})}{\partial x_3}\frac{\partial f_3(\mathbf{x})}{\partial {x_2}}=0
\end{equation}
Where similar equations exists for permutations of the position coordinates. Using the fact that $\frac{\partial \phi(x_{23})}{\partial x_2}=-\frac{\partial \phi(x_{23})}{\partial x_3}$ allows us to write the conditions as:
\begin{align}
\label{eq:third} \frac{\partial f_2(\mathbf{x})}{\partial {x_3}}&=-\frac{\partial f_2(\mathbf{x})}{\partial {x_2}} \\
\frac{\partial f_1(\mathbf{x})}{\partial {x_3}}&=0 \\
\frac{\partial f_3(\mathbf{x})}{\partial {x_2}}&=0
\end{align}

equation \ref{eq:third} will be generically true for a system where the action of the chemical modification is to change a pair potential interaction paramter. The other two conditions are more interesting. It states that in the case that the interaction parameter between particles $i$ and $j$ is changed according to the position of a different particle $k$, the vector field $c$ is no longer conservative. 

\section{Solution of Smoluchowski equation for linear systems}

Starting from the stationarity condition:
\begin{equation} \label{eq:fp}
\nabla.\left(\exp\left(-\phi(\mathbf{r})\right)\left(-\nabla H_0+\mathbf{m}+k_b T  \nabla \phi\right) \right)=0
\end{equation}
 and introducing a new vector $\mathbf{L}=-\nabla H_0+\mathbf{m}+k_b T  \nabla \phi$, leading to $\mathbf{L}-\mathbf{F}=k_b T \nabla \phi$ where $\mathbf{F}$ is the vector of all the forces in the system we obtain the condition that:
\begin{equation}
k_b T \nabla.\mathbf{L}-\mathbf{L}.\left(\mathbf{L}-\mathbf{F}\right)=0
\end{equation}
with the additional condition that $\mathbf{L}-\mathbf{F}$ is conservative. Now we say that $\mathbf{F}=\underline{A} \mathbf{x}$ where $\underline{A}$ is a matrix which does not depend on position and searching for solutions $\mathbf{L}=\underline{B}\mathbf{x}$ we have the following conditions that must be satisfied in order for equation \ref{eq:fp} to hold:
\begin{align}
\text{Tr}(\underline{B})&=0 \label{cond1}\\
\mathbf{x}^T \left(\underline{B}^T \underline{B}-\underline{B}^T \underline{A}\right) \mathbf{x} &=0 \label{cond2}\\ 
(\underline{A}-\underline{B})^T&=\underline{A}-\underline{B} \label{cond3}
\end{align}
condition (\ref{cond2}) is true if the combined matrix is antisymmetric. Upon decomposition of the two matrices $\underline{A},\underline{B}$ into their symmetric and anti symmetric components:
\begin{align}
\underline{A}&=\underline{S}+\underline{Q}\\
\underline{B}&=\underline{\Sigma}+\underline{\Theta}
\end{align}
we can find the elements of matrix $B$ as a function of the given matrix $A$, where condition (\ref{cond3}) immediately leads to $\underline{\Theta}=\underline{Q}$, and we have upon expansion the following equation for $\underline{\Sigma}$
\begin{equation}
2 \underline{\Sigma}^2-\underline{\Sigma}.(\underline{S}+\underline{Q})-(\underline{S}+\underline{Q})^T .\underline{\Sigma}+[\underline{Q},\underline{S}]=0
\end{equation}
where the square brackets denote a commutator. This equation is known as the continuous time algebraic Riccati equation and has well devoloped methods of solution, for example by eigendecomposition of a larger system. We omit full details of how this equation is solved and let the reader consult\cite{lancaster_1995}. Any matrix $\Sigma$ that solves this equation in addition to being symmetric and traceless will be a solution to the Smolochowski equation. We can then compare these solutions to the perturbative expansion and see how well it describes the true solutions to the equations.

\section{Green's functions}

\subsection{Slowly varying approximation}
The Green's function equation in the main text is given by:
\begin{equation} \label{eq:green2}
\left(\nabla^2-\beta \mathbf{\nabla}H_0.\nabla\right)G(\mathbf{r},\mathbf{r'})=\delta(\mathbf{r-r'}) 
\end{equation}
which is a hard problem, as we mentioned. If we assume that the boltzmann weight is slowly varying, i.e., $\nabla^2 \exp(-\beta H_0(\mathbf{r}))\approx 0$ then we can rewrite equation \ref{eq:green2} as:
\begin{equation}
\exp(\beta H_0(\mathbf{r})/2)\nabla^2\left(\exp(-\beta H_0(\mathbf{r})/2)G(\mathbf{r},\mathbf{r'})\right)=\delta(\mathbf{r-r'}) 
\end{equation}
To show this when expanded and non-dimensionalized ($H_0\sim\epsilon \tilde{H}_0$,$\nabla G\sim(1/b)\nabla\tilde{G}$,$\nabla H_0\sim(\epsilon/\lambda)\nabla\tilde{H}_0$) leads to the following equation:
\begin{equation}
\nabla^2 \tilde{G}(\mathbf{r}) - \left(\frac{b \beta \epsilon}{\lambda}\right)\nabla \tilde{H_0}(\mathbf{r}).\nabla \tilde{G}(\mathbf{r})+\frac{\beta \epsilon b^2}{2 \lambda^2}\left(\frac{1}{2}\beta \epsilon \nabla \tilde{H}_0(\mathbf{r})^2-\nabla^2 \tilde{H}_0(\mathbf{r})\right)\tilde{G}(\mathbf{r})=\delta(\mathbf{r-r'}) 
\end{equation}
Therefore the condition under which this equation reduces to equation \ref{eq:green2} is that $b/\lambda\gg(b/\lambda)^2$
Which leads to the solution seen in the main text.

\subsection{Locality of Green's function approximation}

We wish to say how the probability of a state is modified upon the introduction of a non-conservative force. In the main text, the first order perturbation was seen to be:
\begin{equation}
P(\mathbf{r_1})\approx\exp\left(-\beta H_0(\mathbf{x_1})+\beta^2 \int \mathrm{d}\mathbf{r'} G(\mathbf{r}_1,\mathbf{r'}) \rho_s(\mathbf{r'})\right)
\end{equation}
Where $\rho_s(\mathbf{r'})=\nabla H_0(\mathbf{r'}).\mathbf{m}(\mathbf{r'})$ is the source term from the main text.

In order to properly characterize the effect of the non-conservative perturbation, we will compare instead the ratio of probabilities of two different states, as this quantity will be well bounded. We shall also normalize by their probability ratio in the absence of a non-conservative force:
\begin{equation}
P_{\text{eq}}(\mathbf{r_1})=\exp(-\beta H_0(\mathbf{r_1}))
\end{equation}
we therefore introduce the following quantity:
\begin{equation}
R\equiv\log\left(\frac{P(\mathbf{r_1})/P(\mathbf{r_2})}{P_{\text{eq}}(\mathbf{r_1})/P_{\text{eq}}(\mathbf{r_2})}\right)
\end{equation} 
which tells us about the extent to which the non-conservative force has modified the probabilities of states 1 and 2 in excess of their equilibrium probabilites. If this quantity is equal to zero there is no change in the probabilities. Reinserting our expressions for the probabilites above:
\begin{equation}
R=\left(\beta^2 \int \mathrm{d}\mathbf{r'} G(\mathbf{r}_1-\mathbf{r'}) \rho_s(\mathbf{r'})-\beta^2\int \mathrm{d}\mathbf{r'} G(\mathbf{r}_1-\mathbf{r'}) \rho_s(\mathbf{r'}) \right)
\end{equation}
With the assumption that the Green's function is translationally invariant (for example for translationally invariant Hamiltonians), as these are convolutions they can be rewritten as:
\begin{equation}
R=\left(\beta^2 \int \mathrm{d}\mathbf{r'} G(\mathbf{r'}) \rho_s(\mathbf{r_1}-\mathbf{r'})-\beta^2\int \mathrm{d}\mathbf{r'} G(\mathbf{r'}) \rho_s(\mathbf{r}_2-\mathbf{r'}) \right)
\end{equation}
We may Taylor expand the source terms:
\begin{equation}
\rho_s(\mathbf{r_1}-\mathbf{r'})\approx\rho_s(\mathbf{r_1})-\mathbf{r'}.\nabla\rho_s(\mathbf{r_1})+\frac{1}{2}\mathbf{r'}.\underline{H_e}(\mathbf{x_1}).\mathbf{r'}+\hdots
\end{equation}
Where $\underline{H_e}$ is the Hessian matrix. The truncation becomes better the more rapidly $G(\mathbf{r'})$ decays to zero. Additionally, the integral of the odd first derivative term in the expasion with the even Green's function will lead to a zero integral. Therefore if we were only to keep the first few term (corresponding to Green's functions which decays very fast as a function of $\mathbf{x'}$) we are left with the following expression:
\begin{equation}
R\approx\lambda^2 \beta^2 (\rho_s(\mathbf{r_1})-\rho_s(\mathbf{r_2}))+\frac{1}{2}\beta^2\sigma^4(\nabla_1^2 \rho_s(\mathbf{r_1})-\nabla_2^2 \rho_s(\mathbf{r_2}))+\hdots
\end{equation} 
where we introduce the normalization constants such that $\int G(\mathbf{x'})\mathrm{d}\mathbf{x'}=\lambda^2,\int |\mathbf{x'}|^2 G(\mathbf{x'})\mathrm{d}\mathbf{x'}=\sigma^4$ and so on. Therefore, the zeroth order effect from the imposition of non-conservative force fields will be felt as a the local difference between the values of the source terms at different points, as per the main text. Higher order effects will be due to local curvatures of the source at that point, and so on.

\subsection{Effective smearing of source terms}

The full Green's function is very complicated, however from the previous sections here and in the main text we can see effectively it acts as an operator that smears out source terms. In this section we demonstrate qualitatively this effect by solving numerically the equation:

\begin{equation} \label{eq:perturb}
\left(\nabla^2-\beta \nabla H_0(\mathbf{r}).\nabla\right)\mu_1(\mathbf{r})=\beta^2 (\nabla H_0(\mathbf{r})).\mathbf{m}(\mathbf{r})
\end{equation}
and comparing the solution $\mu_1$ to the source term $\beta^2 (\nabla H_0(\mathbf{r})).\mathbf{m}(\mathbf{r})$

\begin{figure}[!h]
\begin{center}
\includegraphics[width=180mm]{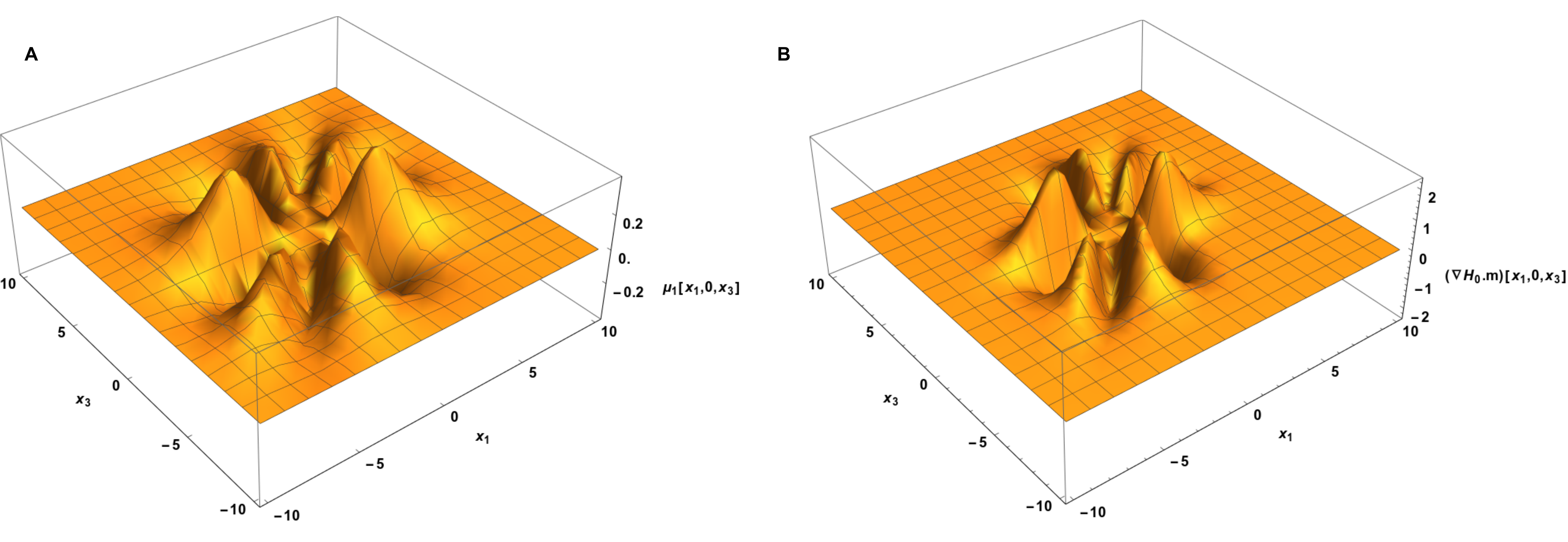}
\caption{Comparison of the value of the solution (A)  with the source term (B) for the $x_2=0$ slice. The features of the solution are qualitatively similar to the source term, slightly smeared and rescaled. }
\label{fig:sol} %
\end{center}
\end{figure}

We choose a three dimensional system $(\mathbf{r})=(x_1,x_2,x_3)$ with the following Hamiltonian:
\begin{equation}
H=\exp(-\sum_i x_i^2)+\sum_{i>j} \exp(-(x_i-x_j)^2/2\sigma^2)
\end{equation}
with the following choice for $\mathbf{m}$
\begin{equation}
\mathbf{m}=\nabla\times(1,1,1)\sum_{i>j}\exp(-(x_i-x_j)^2/2\lambda^2)
\end{equation}
We show the solution and the source terms in figure  \ref{fig:sol}

\bibliographystyle{unsrt}
\bibliography{exported-references}